\documentclass[aps,pra, twocolumn,superscriptaddress]{revtex4-2}
\usepackage{graphicx}% Include figure files
\usepackage{dcolumn}% Align table columns on decimal point
\usepackage{bm }% bold math
\usepackage{paralist}
\usepackage{braket}
\usepackage{hyperref}
\usepackage{natbib}
\usepackage{xcolor}
\usepackage{upgreek}
\usepackage{comment}
\usepackage{makecell}

\newcommand{\Ca}{\ensuremath{^{40}{\rm Ca}^+\,}}
\newcommand{\mum}{$\mu$m}

\begin{document}

\title{Scalable Ion Fluorescence Collection Using a Trap-Integrated Metalens}

\author{Hae Lim}
\affiliation{Department of Electrical and Computer Engineering, The University of Washington, Seattle WA 98103, USA}
\author{Johannes E. Fr{\"o}ch}
\affiliation{Department of Electrical and Computer Engineering, The University of Washington, Seattle WA 98103, USA}
\author{Christian M. Pluchar}
\affiliation{Department of Electrical and Computer Engineering, The University of Washington, Seattle WA 98103, USA}
\author{Arka Majumdar}
\affiliation{Department of Electrical and Computer Engineering, The University of Washington, Seattle WA 98103, USA}
\affiliation{Department of Physics, The University of Washington, Seattle WA 98103, USA}
\author{Sara L. Mouradian}
\affiliation{Department of Electrical and Computer Engineering, The University of Washington, Seattle WA 98103, USA}

\begin{abstract}
A scaled trapped-ion quantum computer will require efficient fluorescence collection across a large area. Here we propose and demonstrate a compact monolithically integrated system featuring a metalens fabricated on the backside of a surface ion trap. A 40$\times100$\,\mum{} aperture enables a simulated point-source collection efficiency of 0.91\% and a measured point-source detection efficiency of 0.58\%. Increasing the aperture area to 40$\times600$\,\mum{} boosts the simulated collection efficiency to 3.17\%—comparable to that of a conventional objective with a numerical aperture of 0.35. Further improvements are possible by co-optimizing the electrode and aperture geometry. An undercut of the electrode substrate at the aperture ensures a large distance between the ion and dielectric substrate without compromising collection efficiency. The metalens directly collimates the collected fluorescence, eliminating the need for a high numerical aperture objective. %We find that the detection efficiency does not decrease substantially over a 1\,cm diameter area. 
An array of such readout zones will offer a compact, scalable solution for high-fidelity parallel readout in next-generation trapped-ion quantum processors.

\end{abstract}

\maketitle

Quantum computers capable of demonstrating practical quantum advantage are projected to require hundreds of logical qubits~\cite{doi:10.1126/sciadv.1601540, doi:10.1073/pnas.1619152114}, corresponding to millions of physical qubits with current error correction codes~\cite{PhysRevResearch.2.033128}. Trapped ions are a leading platform for quantum computation~\cite{Bruzewicz2019Jun, 9797817, PhysRevX.14.041017}, but even the largest platforms~\cite{Chen_2024,DeCross_2025} are currently limited to tens of ions. It is thus necessary to build systems that support scaling to millions of physical qubits while preserving high fidelity initialization, gate execution, and state discrimination. Here, we consider the quantum charge coupled device (QCCD) architecture~\cite{Kielpinski2002Jun} where millions of ions will be trapped in short linear chains and shuttled between initialization, control, and readout zones on a cm-scale chip. In particular, we focus on achieving high fidelity, fast state discrimination of trapped-ion qubits which relies on efficient fluorescence detection. Achieving scaled, high-fidelity state discrimination thus necessitates the development of advanced optical systems capable of high collection efficiency and uniform performance across multiple readout zones on a cm-sale device.

\begin{figure}
    \centering
    \includegraphics[width=0.95\linewidth]{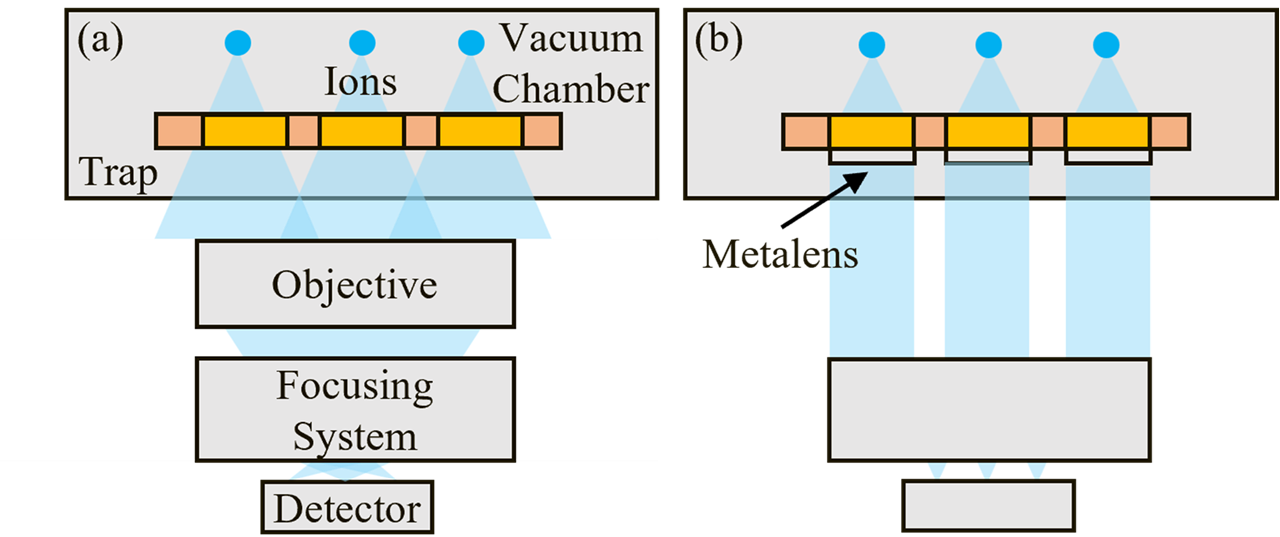}% Draft/intro.png}
    \caption{(a) Conventional objective-based optical system used to image trapped-ion fluorescence (not to scale). (b) Scalable fluorescence collection with collimating metalenses integrated at each readout zone (not to scale).}
    \label{intro}
\end{figure}

Currently, most trapped-ion quantum computing demonstrations use a single objective to collect ion fluorescence as depicted in Fig.~\ref{intro}\,(a). While it may be possible to design an objective that provides high collection efficiency over a cm-scale field of view, such as the objectives used for photolithography~\cite{Zheng2021Mar}, this is likely to be bulky and difficult to integrate with the full trapped-ion apparatus.
%However, designing and building a single high numerical aperture (NA) objective lens with a cm-scale field of view would require difficult and expensive optical engineering. 
Thankfully, high numerical aperture (NA) readout is only necessary in dedicated readout zones in a QCCD architecture,, and not continuously across the chip as in a conventional imaging system. Previous works have exploited this to create dedicated detection zones with diffractive mirrors~\cite{Ghadimi_2017, Connell_2021}, detectors~\cite{PhysRevLett.126.010501,PhysRevLett.129.100502,Setzer2021Oct} or collection-optimized grating couplers~\cite{Knollmann2024Aug,Smedley2025Feb} directly integrated with the surface ion trap electrodes, achieving collection efficiencies of a few percent. %Unfortunately, these solutions can cause increased motional heating and also preclude post-collection spatial filtering to reduce background noise. 

Here, we take a different approach and design a system with small high-NA lenses directly integrated with the readout zones of a surface ion trap, as pictured in Fig.~\ref{intro}\,(b). These integrated lenses collimate the ion fluorescence at each zone. This system can be paired with a focusing system and detector for state detection. Our device decouples the field of view and NA of the system, enabling a readout area limited only by the number of readout zones and the size of the focusing system and detector. We design, fabricate, and test one unit cell of such a device, demonstrating that it provides the same collection efficiency as a conventional objective, but with an expanded field of view limited only by the size of the focusing system and detector. 

Previous efforts have mechanically integrated lenses with surface ion traps~\cite{Brady_2011, PhysRevLett.106.010502, PhysRevApplied.1.024004}. In contrast, our approach involves fabricating a lens, specifically a metalens, directly on the backside of the trap itself.
%In conventional imaging systems, such as the two-lens setup illustrated in Fig.~\ref{intro} (a), the collection efficiency of fluorescence emitted by trapped ions diminishes as the readout zone position moves farther from the optical axis of the lenses. This reduction in efficiency is primarily due to geometric and optical limitations that restrict the effective collection angle for off-axis points. One possible solution is to increase the size of the optical components—such as using lenses with larger slot—to improve light collection across a wider field. However, this approach further enlarges an optical system that is already bulky.%, making it less suitable for scalable architectures involving multiple readout zones.
%An alternative approach involves miniaturizing the collimating lens while preserving its numerical aperture (NA), which is crucial for maintaining efficient light collection from the ion. By reducing the physical dimensions of the lens without compromising its optical performance, it can be directly integrated with a surface ion trap, positioned close to the ion to collect its emitted fluorescence. 
Metalenses enable direct integration and provide a smaller footprint and lower volume than refractive lenses~\cite{engelberg2020advantages,kuznetsov2024roadmap,shen2021chip,huang2019monolithic}. This class of diffractive optical elements consists of sub-wavelength scale scatterers. Each scatterer imparts a specific phase delay, based on its physical dimension~\cite{arbabi2015subwavelength}. Thus, arbitrary phase profiles, such as a high-NA focusing profile\cite{paniagua2018metalens}, extended depth of focus\cite{huang2020design}, or polarization-multiplexed focusing \cite{froch2025full}, can be implemented in a single layer of a dielectric material. A single metalens device can also be designed to have different functions at different wavelengths, which are necessary for the optical control of atomic qubits~\cite{Park2024Sep}. Moreover, metalens fabrication relies on standard cleanroom or foundry processes~\cite{yang2025nanofabrication} and can thus be integrated with high accuracy with respect to the trap. This delivers a single chip solution with both trap and optical elements, making post-fabrication alignment obsolete~\cite{Hu2022Design, PRXQuantum.3.030316, Ropp2023Integrating, holman2024trappingsingleatomsmetasurface, chen2024multifunctionalmetalenstrappingcharacterizing}.

%A further critical advantage over classical refractive lenses, is illustrated in Fig.~\ref{intro} (b), highlighting how this compact lens system can be duplicated at each readout zone across the trap, enabling localized fluorescence collection at every zone. The collimated light from each zone can then be directed into a shared, global optical system that relays the fluorescence onto a camera or an array of single photon detectors, such as an electron multiplying charge-coupled device (EMCCD). This architecture supports scalable, parallel ion readout while minimizing alignment complexity. 

\begin{figure}
    \centering
    \includegraphics[width=0.95\linewidth]{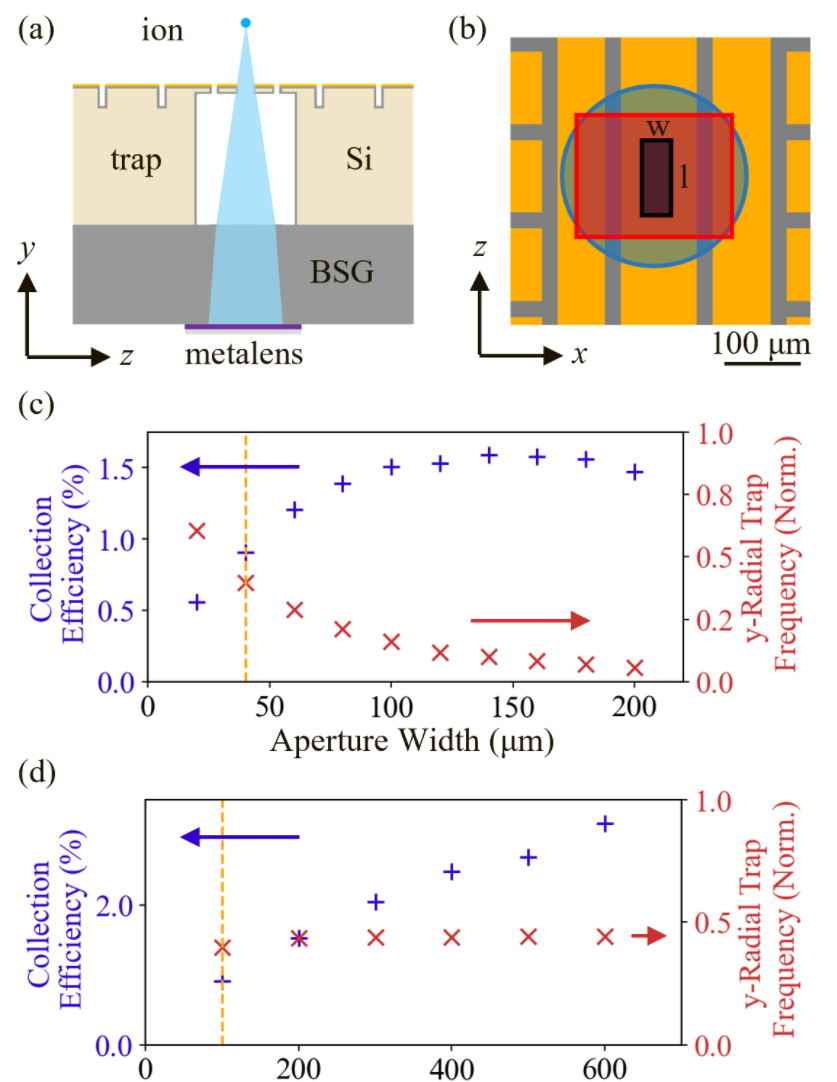}
    \caption{(a) Cross-section view of the device (not to scale). (b) Top view of the device, showing the alignment and relative sizes of the aperture (black), undercut (red), and metalens (black). (c) Simulated collection efficiency and normalized $y$-radial trap frequency as function of aperture width. The yellow dashed line indicates the aperture width of the fabricated device. (d) Simulated collection efficiency and normalized $y$-radial trap frequency with the undercut as function of aperture length. The yellow dashed line indicates the aperture length of the fabricated device.}
    \label{device_sim}
\end{figure}

We begin with a standard surface trap design~\cite{Chiaverini_2005} and modify it to enable fluorescence collection through the trap. This enables direct integration with fabrication-limited alignment tolerances and will be directly applicable to multi-wafer trap designs~\cite{wilpers-2012, 10.1063/1.5119785, Decaroli_2021} which may be necessary in a large-scale architecture~\cite{Nguyen2025May}. 
To facilitate light collection, an aperture is patterned into the ground electrode, allowing a fraction of the ion’s fluorescence to pass through the trap and reach the integrated lens. This backside lens collimates the transmitted fluorescence, which is then relayed to a global focusing system and ultimately onto a detector. For our analysis, we assume the ion emits photons uniformly in all directions, though note that the actual collection efficiency would depend on the dipole orientation. We define the collection efficiency as the percentage of total emitted light that successfully passes through the trap and is incident on the lens. Detection efficiency, on the other hand, accounts for additional losses along the optical path to the detector.

Overhanging electrodes provide a means to increase the distance between the trap electrodes and the underlying dielectric substrate while preserving the optical collection efficiency of the system. This design should mitigate motional heating of the trapped ion due to dielectric surfaces~\cite{PhysRevLett.126.230505}. The structural configuration of our device is illustrated in Fig.~\ref{device_sim} (a), where the undercut region beneath the electrodes and the edges of the integrated metalens are marked in red and blue, respectively. The size of the undercut and lens should be chosen such that the aperture defines the collection efficiency.

To determine the optimal parameters for our device, we simulate the electric potential generated by the trap electrodes using COMSOL Multiphysics. From the resulting static potential, we compute the pseudopotential~\cite{Littich} to obtain the height of the radio frequency (rf) null and the radial trap frequencies for a given rf drive voltage and frequency. For all simulations, we fix the dimension of the rf electrodes to 65\,\mum{} and 80\,\mum, with an inter-electrode spacing of 20\,\mum. The aperture is introduced within the ground electrode, which has a baseline width of 65\,\mum.

We investigate the effect of aperture dimension on the collection efficiency. An increased aperture width increases the overall size of the ground electrode and thus the distance between the rf electrodes, increasing trap height. Fig.~\ref{device_sim}\,(c) illustrates the effect of varying the aperture width (aperture length = 100\,\mum) on collection efficiency and the $y$-radial trap frequency, normalized to the trap frequency without an aperture. While widening the aperture would increase the collection efficiency for a given trap height, the trap height also increases with aperture size leading to an optimal collection efficiency near an aperture width of 150\,\mum. %This behavior arises from the interplay between the increasing aperture size and the corresponding trap height. Initially, as the aperture widens, its growth outpaces the increase in trap height, leading to improved collection of fluorescence due to a larger acceptance angle. However, beyond this point, the trap height begins to grow more rapidly than the aperture width. This relative geometric change reduces the effective solid angle subtended by the aperture, thereby causing a gradual decline—or roll-off—in collection efficiency. 
Taking into account both optical efficiency and practical rf voltage requirements, we select a conservative aperture width of 40\,\mum{}. 

As seen in Fig.~\ref{device_sim}\,(d), increasing the length of the aperture has nearly no effect on the trap frequency and leads to a marked increase in collection efficiency.
% As the slot width increases, the two ground electrodes on either side of the slot are held constant at 32.5\,$\mu$m each. The slot length is fixed at 100\,$\mu$m, though further investigation is needed to determine whether this constraint is optimal.
%As the slot widens, the adjacent rf electrodes are displaced outward to accommodate the geometry. This increases the rf null height above the electrodes, which in turn weakens the confinement and reduces the radial trap frequencies. From an optical standpoint, a wider slot increases the solid angle available for collecting fluorescence, thereby improving collection efficiency. However, this benefit plateaus and eventually reverses: beyond a slot width of approximately 140\,$\mu$m, the detrimental effect of the increased rf null height outweighs the optical gain, leading to reduced overall performance.
We choose a 100\,\mum{} aperture length to ensure structural stability, which yields a simulated fluorescence collection efficiency of 0.91\%. For this geometry, with an rf drive voltage of 50\,V at 20\,MHz, we expect a $x$ ($y$) radial trap frequency of 1.21\,MHz (1.36\,MHz) for a \Ca ion. Increasing the aperture length to 600\,\mum{} improves the simulated collection efficiency to 3.17\,\%, comparable to that of a conventional objective with a NA of 0.35. Without an undercut region, however, the optical path would be blocked by the bulk material, reducing the simulated collection efficiency to just 0.20\% for a 275\,\mum{} height difference between the trap electrode and dielectric surface, as in our device.

%By introducing an undercut to the substrate, this obstruction is removed, significantly increasing the accessible solid angle for light collection. Our simulated performance is comparable to recent results achieved with monolithically integrated detectors in surface traps ~\cite{PhysRevLett.126.010501, PhysRevLett.129.100502}, and we anticipate that further improvements in collection efficiency are achievable through continued optimization of the electrode geometry and integrated optical components.

After settling on a final trap design, we optimized the metalens to collimate the 397\,nm fluorescence from a \Ca ion trapped at the rf null above the center of the aperture. We used a flat phase profile in Zemax Lightstudio, considering the dimensions of the trap and the thickness of the glass on which the metalens is fabricated. Assuming a radial symmetric Zernike polynomial with even summands, we optimized the focus of collimated light at the trap location. This phase profile was then converted to a feature map, based on a precalculated library of phase delay vs. feature size for 700\,nm tall silicon nitride (SiN) pillars with a period of 250\,nm \cite{zhan2016low}.

\begin{figure}
    \centering
    \includegraphics[width=1\linewidth]{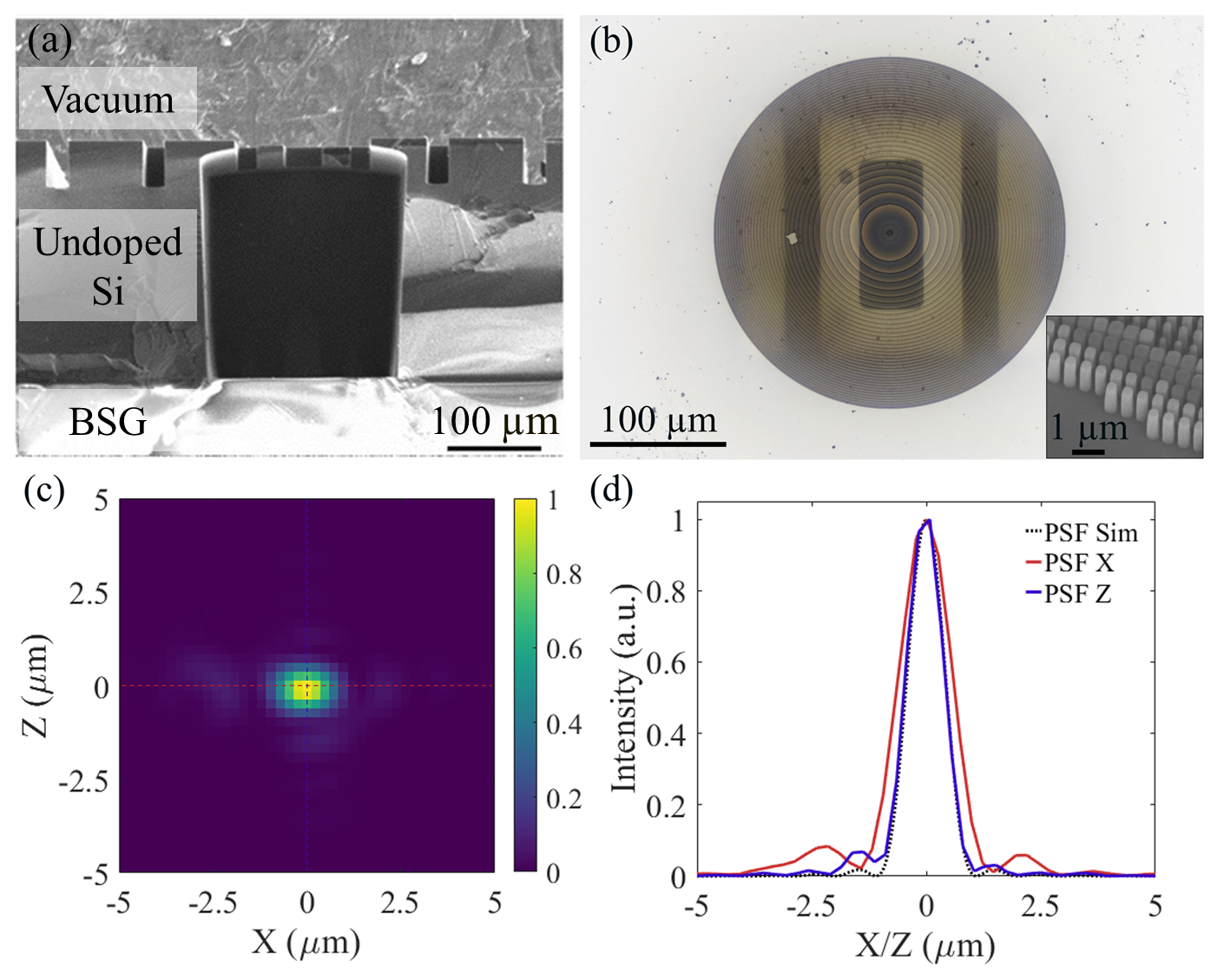}
    \caption{(a) SEM image showing a cross-sectional view of the fabricated device near the center of the aperture, after bonding the silicon and BSG substrates. (b) Overlayed optical images displaying the alignment between the surface trap, the undercut region in the silicon substrate, and the metalens fabricated on the BSG substrate, viewed through the chip. The inset shows a SEM image of the metalens in oblique view. (c) Transverse intensity profile of the 397\,nm laser beam at the focal plane, showing the beam’s spatial distribution and focal spot size. (d) Simulated PSF and experimentally measured PSF along the $x$ and $z$ directions.}
    \label{fab_char}
\end{figure}

There are three components of the device that must be lithographically aligned - the trenches defining the electrodes, the undercut region to maintain collection efficiency, and the metalens. The electrodes and undercut are etched into a 275\,\mum{} thick undoped Si substrate and the metalens is defined in a 700\,nm SiN on a 200\,\mum{} thick borosilicate glass (BSG) substrate. We begin with a deep reactive ion etch into an undoped silicon substrate to define the undercut and backside alignment marks. We then bond the etched side to the BSG substrate. Using the backside alignment marks we pattern and etch the trenches to define the electrodes, the aperture, and alignment marks on the top unbonded side of the silicon substrate. A cross-sectional scanning electron microscope (SEM) image of the aperture region of the device after etching the trenches is shown in Fig.~\ref{fab_char}\,(a). Thermal oxidation and Cr and Au evaporation complete the electrodes. 

Next, we deposit a thin film of SiN using plasma enhanced chemical vapor on the unbonded side of the BSG substrate and fabricate markers, establishing alignment features for the metalens fabrication. Then we spin coat a resist layer on top of the backside of the chip and pattern the aforementioned optimized feature map into that resist layer using electron beam lithography. Following development of the resist, electron beam evaporation is used to deposit a 70\,nm thick layer of alumina, which is then lift off, defining the hard mask. Finally, we used reactive ion etching to transfer the pattern from the hard mask into the underlying SiN thin film~\cite{zhan2016low}. Fig.~\ref{fab_char}\,(b) shows an overlay of images of the integrated trap (located on the top side of the chip) and metalens (located on the backside of the chip). The inset shows an SEM image of the metalens, highlighting the structural integrity of the nanopillars after fabrication. %During fabrication, we correlated the relative position of the metalens to the top of the trap such that aligned fabrication is within micron accuracy.

To characterize the properties of the integrated device, we measured the point spread function (PSF) formed under illumination with a collimated laser at 397\,nm. With light incident on the metalens from the backside of the trap we measured the PSF on the top side of the sample using a microscope mounted on a programmable translation stage. Fig.~\ref{fab_char}\,(c) shows a measurement of the the PSF at a distance of 125\,\mum{} relative to the trap surface, the target location of the ion. A direct comparison of the measured PSF profile with the simulated profile is shown in Fig.~\ref{fab_char}\,(d). We observe that the full-width half-maximum (FWHM) of the $z$ profile (0.95\,\mum) matches closely with the FWHM of the simulated PSF profile (0.92\,\mum), whereas the $x$ profile, shows a slightly larger FWHM (1.31\,\mum), due to the aperture width. The experimentally observed profiles also show slightly larger intensity in their sidelobes, which may be related to fabrication imperfection. %Nevertheless, this shows that clear focal spots at the ion trap location can be achieved, which are within suitable limits.  
\begin{figure*}[hbtp]
    \centering
    \includegraphics[width=0.95\linewidth]{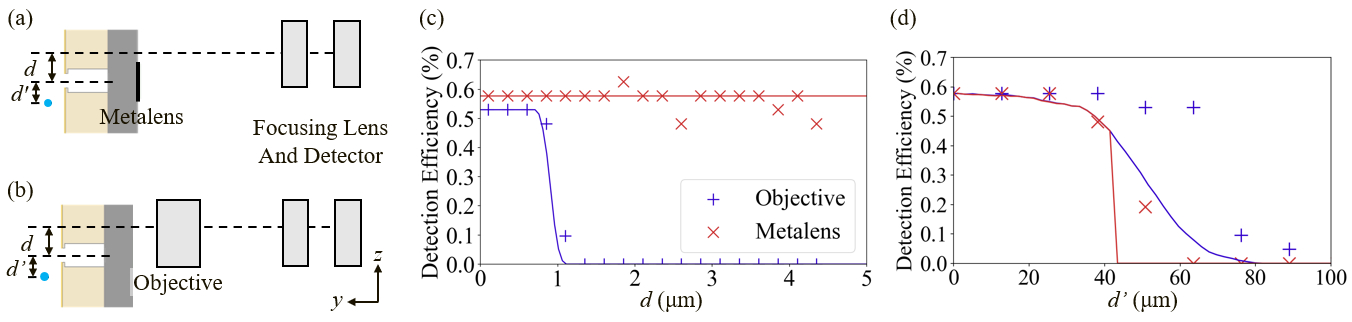}
    \caption{(a) Schematic of the integrated setup, where a collimating metalens is monolithically integrated beneath the trap to direct fluorescence parallel to the optical axis (not to scale). (b) Schematic of the objectivec setup, in which a surface trap is used without an integrated collimating lens (not to scale). (c) Experimental measurements of detection efficiencies for both setups as functions of the lateral displacement $d$ between the readout zone, with the ion held above the center of the aperture at the rf null, and the optical axis of the rest of the system. Solid lines correspond to simulation results obtained using Zemax OpticsStudio, with an ideal thin lens substituted in place of the metalens to model collection behavior under idealized conditions. (d) Detection efficiencies in both setups measured as functions of the axial displacement $d'$ of the ion from the center of the aperture at the rf null along the lateral (z) direction, representing misalignment of the ion’s axial position relative to the optical focus.}
    \label{meas}
\end{figure*}

After validating the optical performance of a single device, we examine the scalability of our system. Our claim is that collimation optics local to each readout zone (Fig.~\ref{intro}(b)) will provide a larger field of view in comparison to a system with a single objective (Fig.~\ref{intro}(a)) for a given imaging system. We compare our system to a system with objective with NA = 0.5 (Nikon, MRH00205), higher than the NA of the aperture. Instead of fabricating multiple readout zones, we measure the effect of readout zone position by translating the device with respect to the free-space optics ($d$) as illustrated in Fig.~\ref{meas}(a,b). For both systems the collimated light is focused onto a detector using a lens with a 1 inch diameter.

We create an ``artificial ion'' by tightly focusing a 397\,nm laser at the expected trap height above the center of the
aperture, producing a beam with a measured divergence angle of 10.98  $\pm$ 1.53$^{\circ}$, which exceeds the 8.46$^{\circ}$ acceptance angle of the aperture. At each displacement $d$ (as shown in Fig.~\ref{meas}\,(a,b)), we measure the power at the detector and normalize it to the optical power incident on the trap. Using the measured divergence angle, we calculate the expected fluorescence detection efficiency assuming isotropic emission. Fig.~\ref{meas}\,(c) plots the results for the single objective (blue) and integrated device array (red). We compare our measurements to ray-based simulations performed in Zemax OpticsStudio (solid lines), normalized to the measured detection efficiency.

In the metalens setup at $d$ = 0\,mm, our simulation predicts a collection efficiency of 0.91\%, whereas the measured detection efficiency is 0.58\%. This discrepancy primarily arises from the 67\% transmittance through the BSG and metalens. The remaining discrepancy likely arises from imperfect collimation of the metalens and the finite size of the focusing lens and detector.

%To facilitate comparison, the simulated detection efficiencies were normalized to the measured efficiency at $d$ = 0\,mm. For example, in the simulation of the objective setup, the detection efficiency at $d$ = 0\,mm was 0.77\%. Despite idealized components in the simulation, this is lower than the 0.91\% collection efficiency predicted by a COMSOL model. This discrepancy arises from differences in source modeling: the Zemax simulations used a Gaussian beam, with 86.5\% of the total power confined within the 10.98$^{\circ}$ divergence angle, while the COMSOL model assumed an ideal isotropic emitter with 100\% of the power distributed within a comparable angle. Ultimately, all simulated detection efficiencies, including the 0.77\% value at $d$ = 0\,mm, were scaled to match the measured value of 0.58\% at $d$ = 0\,mm.

The limited field of view of the single-objective system is evident in Fig.~\ref{meas}(c). The detection efficiency begins to decrease sharply at a distance of approximately 0.80\,mm, reaching 0\% at a displacement of 1.2\,mm. This is consistent with the ray-based simulations (solid line) which predict a drop to 90\% of the maximum achievable detection efficiency at 0.86\,mm. %This shows that light from the readout zones off the optical access will suffer from reduced detection efficiency in a single-objective configuration. We compare our measurements to ray-based simulations performed in Zemax OpticsStudio (solid lines). In the simulations, all objects were set to be ideal except that 

In contrast, the integrated setup shows only a gradual decline in detection efficiency as the device is moved off-axis. This broader effective collection range is because the collimating metalens is directly integrated at each readout zone. In the ray-based optics simulation (solid line), the detection efficiency does not drop until 12.63\,mm, the radius of the focusing lens. However, in practice the beam is not perfectly collimated by the metalens and less of the beam will be incident on the focusing lens for readout zones further from the opitcal access, reducing detection efficiency. This comparison highlights the advantage of integrating the collimating lens directly with the trap for scalable ion trap architectures.

Finally, we investigate the performance of the two setups for the measurement of many ions in a linear chain. To do this, we fixed the position of the device and imaging system in both setups, translated the position of the artificial ion source along the axial ($z$) direction ($d'$ in Fig.~\ref{meas}\,(a,b)), and measured the optical power collected after the imaging system. %This measurement is particularly relevant to practical implementations of trapped-ion systems, where ions are often arranged in a linear chain rather than positioned directly at the rf null. As a result, understanding how detection efficiency varies with the ion’s axial position is crucial for assessing the robustness of fluorescence collection across a realistic spatial range within a readout zone.

The measured detection efficiency as a function of axial ion displacement ($d'$) is plotted in Fig.~\ref{meas}\,(d) along with the simulated detection efficiency. In both setups detection efficiency decreases as the ion is moved from the center of the aperture as the light is blocked by the trap electrodes. The integrated metalens shows slightly worse performance as it was optimized to collimate light from a single source directly on the optical access. In future iterations, the metalens could be specifically designed to achieve focusing across a chain of ions.%farther from the rf null. This trend occurs because, with increasing axial displacement, more of the emitted fluorescence becomes occluded by the trap resulting in less light being transmitted through to the collection optics. In the conventional setup, the detection efficiency gradually drops to zero at a displacement of 64\,\mum. Beyond this point, the fluorescence beam is fully blocked by the trap and no light reaches the focusing lens. In the integrated setup, the detection efficiency reaches zero at a smaller displacement of 38\,\mum. This sharper cutoff arises because the collimating metalens is positioned directly beneath the ion and is designed to operate optimally only when the ion is located near its focus or the rf null. As the ion moves farther from the rf null, the metalens is no longer able to properly collimate the emitted fluorescence, and the resulting beam diverges away from the optical axis, missing the focusing lens entirely.

%The reduced axial collection range of the integrated system can be compensated for by increasing the slot size of the focusing lens. This adjustment allows for a broader acceptance of light from off-focal positions without requiring any increase in detector size. In contrast, scaling the detection system by enlarging the camera presents a more substantial challenge.

This work demonstrates a scalable and efficient approach to fluorescence collection for trapped-ion quantum computing through the monolithic integration of a collimating metalens with a surface ion trap. By incorporating an undercut beneath the trap electrodes, we achieve high collection efficiency. Simulation and experimental results confirm that this integrated system will provide a high collection efficiency over a large area, limited only by the number of readout zones and the size of the focusing system and detector. %maintains superior lateral detection range compared to conventional systems, supporting parallel readout over a broader spatial extent. Although the axial collection tolerance within a readout zone is narrower due to the metalens’s focal constraints, this limitation can be readily mitigated with modest optical adjustments. 
Overall, our integrated design offers a compact solution well-suited to the demands of large-scale, high-fidelity ion trap quantum processors and highlights the promise of metalenses for trapped-ion quantum information processing. %Future enhancements in metasurface design and electrode geometry optimization are expected to further improve performance, making this architecture a strong candidate for next-generation scalable quantum systems.
 
We thank Dr. Brant Bowers for his help at the design stage of this project. We also gratefully acknowledge support from the National Science Foundation (NSF) through award ECCS-2240229, as well as funding from the Advancing Quantum-Enabled Technologies (AQET) traineeship program at the University of Washington, which supports interdisciplinary research and training in quantum information science and engineering. A portion of this work was carried out at the Washington Nanofabrication Facility and the Molecular Analysis Facility, which are part of the National Nanotechnology Coordinated Infrastructure (NNCI) at the University of Washington. These facilities receive partial support from the NSF under awards NNCI-1542101 and NNCI-2025489. Additionally, this material is based upon work supported by the National Science Foundation Graduate Research Fellowship Program under Grant No. DGE-2140004.

\bibliography{TrappedIonMetalens}

\end{document}